\documentclass[twocolumn,letterpaper,aps,prl,superscriptaddress,showpacs,floatfix]{revtex4-1}

\usepackage{graphicx}	
\usepackage{xspace}	

\newcommand{\sqsn}{\mbox{$\sqrt{s_{_{NN}}}$}\xspace}

\begin{document}

\title{Beam-energy and system-size dependence of the space-time extent of 
the pion emission source produced in heavy ion collisions}

\newcommand{\abilene}{Abilene Christian University, Abilene, Texas 79699, USA}
\newcommand{\augie}{Department of Physics, Augustana College, Sioux Falls, South Dakota 57197, USA}
\newcommand{\banaras}{Department of Physics, Banaras Hindu University, Varanasi 221005, India}
\newcommand{\barc}{Bhabha Atomic Research Centre, Bombay 400 085, India}
\newcommand{\baruch}{Baruch College, City University of New York, New York, New York, 10010 USA}
\newcommand{\bnlcoll}{Collider-Accelerator Department, Brookhaven National Laboratory, Upton, New York 11973-5000, USA}
\newcommand{\bnlphys}{Physics Department, Brookhaven National Laboratory, Upton, New York 11973-5000, USA}
\newcommand{\caucr}{University of California - Riverside, Riverside, California 92521, USA}
\newcommand{\charlesczech}{Charles University, Ovocn\'{y} trh 5, Praha 1, 116 36, Prague, Czech Republic}
\newcommand{\chonbuk}{Chonbuk National University, Jeonju, 561-756, Korea}
\newcommand{\ciae}{Science and Technology on Nuclear Data Laboratory, China Institute of Atomic Energy, Beijing 102413, People's~Republic~of~China}
\newcommand{\cns}{Center for Nuclear Study, Graduate School of Science, University of Tokyo, 7-3-1 Hongo, Bunkyo, Tokyo 113-0033, Japan}
\newcommand{\colorado}{University of Colorado, Boulder, Colorado 80309, USA}
\newcommand{\columbia}{Columbia University, New York, New York 10027 and Nevis Laboratories, Irvington, New York 10533, USA}
\newcommand{\czechtech}{Czech Technical University, Zikova 4, 166 36 Prague 6, Czech Republic}
\newcommand{\dapnia}{Dapnia, CEA Saclay, F-91191, Gif-sur-Yvette, France}
\newcommand{\debrecen}{Debrecen University, H-4010 Debrecen, Egyetem t{\'e}r 1, Hungary}
\newcommand{\elte}{ELTE, E{\"o}tv{\"o}s Lor{\'a}nd University, H-1117 Budapest, P\'azmany P\'eter s\'et\'any 1/A, Hungary}
\newcommand{\ewha}{Ewha Womans University, Seoul 120-750, Korea}
\newcommand{\fit}{Florida Institute of Technology, Melbourne, Florida 32901, USA}
\newcommand{\fsu}{Florida State University, Tallahassee, Florida 32306, USA}
\newcommand{\gsu}{Georgia State University, Atlanta, Georgia 30303, USA}
\newcommand{\hanyang}{Hanyang University, Seoul 133-792, Korea}
\newcommand{\hiroshima}{Hiroshima University, Kagamiyama, Higashi-Hiroshima 739-8526, Japan}
\newcommand{\howard}{Department of Physics and Astronomy, Howard University, Washington, DC 20059, USA}
\newcommand{\ihepprot}{IHEP Protvino, State Research Center of Russian Federation, Institute for High Energy Physics, Protvino, 142281, Russia}
\newcommand{\illuiuc}{University of Illinois at Urbana-Champaign, Urbana, Illinois 61801, USA}
\newcommand{\inrras}{Institute for Nuclear Research of the Russian Academy of Sciences, prospekt 60-letiya Oktyabrya 7a, Moscow 117312, Russia}
\newcommand{\instpasczech}{Institute of Physics, Academy of Sciences of the Czech Republic, Na Slovance 2, 182 21 Prague 8, Czech Republic}
\newcommand{\isu}{Iowa State University, Ames, Iowa 50011, USA}
\newcommand{\jaea}{Advanced Science Research Center, Japan Atomic Energy Agency, 2-4 Shirakata Shirane, Tokai-mura, Naka-gun, Ibaraki-ken 319-1195, Japan}
\newcommand{\jinrdubna}{Joint Institute for Nuclear Research, 141980 Dubna, Moscow Region, Russia}
\newcommand{\jyvaskyla}{Helsinki Institute of Physics and University of Jyv{\"a}skyl{\"a}, P.O.Box 35, FI-40014 Jyv{\"a}skyl{\"a}, Finland}
\newcommand{\kek}{KEK, High Energy Accelerator Research Organization, Tsukuba, Ibaraki 305-0801, Japan}
\newcommand{\korea}{Korea University, Seoul, 136-701, Korea}
\newcommand{\kurchatov}{Russian Research Center ``Kurchatov Institute," Moscow, 123098 Russia}
\newcommand{\kyoto}{Kyoto University, Kyoto 606-8502, Japan}
\newcommand{\labllr}{Laboratoire Leprince-Ringuet, Ecole Polytechnique, CNRS-IN2P3, Route de Saclay, F-91128, Palaiseau, France}
\newcommand{\lahorelums}{Physics Department, Lahore University of Management Sciences, Lahore 54792, Pakistan}
\newcommand{\lawllnl}{Lawrence Livermore National Laboratory, Livermore, California 94550, USA}
\newcommand{\losalamos}{Los Alamos National Laboratory, Los Alamos, New Mexico 87545, USA}
\newcommand{\lpc}{LPC, Universit{\'e} Blaise Pascal, CNRS-IN2P3, Clermont-Fd, 63177 Aubiere Cedex, France}
\newcommand{\lund}{Department of Physics, Lund University, Box 118, SE-221 00 Lund, Sweden}
\newcommand{\maryland}{University of Maryland, College Park, Maryland 20742, USA}
\newcommand{\mass}{Department of Physics, University of Massachusetts, Amherst, Massachusetts 01003-9337, USA }
\newcommand{\michigan}{Department of Physics, University of Michigan, Ann Arbor, Michigan 48109-1040, USA}
\newcommand{\muenster}{Institut f\"ur Kernphysik, University of Muenster, D-48149 Muenster, Germany}
\newcommand{\muhlenberg}{Muhlenberg College, Allentown, Pennsylvania 18104-5586, USA}
\newcommand{\myongji}{Myongji University, Yongin, Kyonggido 449-728, Korea}
\newcommand{\nagasaki}{Nagasaki Institute of Applied Science, Nagasaki-shi, Nagasaki 851-0193, Japan}
\newcommand{\newmex}{University of New Mexico, Albuquerque, New Mexico 87131, USA }
\newcommand{\nmsu}{New Mexico State University, Las Cruces, New Mexico 88003, USA}
\newcommand{\ohio}{Department of Physics and Astronomy, Ohio University, Athens, Ohio 45701, USA}
\newcommand{\ornl}{Oak Ridge National Laboratory, Oak Ridge, Tennessee 37831, USA}
\newcommand{\orsay}{IPN-Orsay, Universite Paris Sud, CNRS-IN2P3, BP1, F-91406, Orsay, France}
\newcommand{\peking}{Peking University, Beijing 100871, People's~Republic~of~China}
\newcommand{\pnpi}{PNPI, Petersburg Nuclear Physics Institute, Gatchina, Leningrad Region, 188300, Russia}
\newcommand{\riken}{RIKEN Nishina Center for Accelerator-Based Science, Wako, Saitama 351-0198, Japan}
\newcommand{\rikjrbrc}{RIKEN BNL Research Center, Brookhaven National Laboratory, Upton, New York 11973-5000, USA}
\newcommand{\rikkyo}{Physics Department, Rikkyo University, 3-34-1 Nishi-Ikebukuro, Toshima, Tokyo 171-8501, Japan}
\newcommand{\saispbstu}{Saint Petersburg State Polytechnic University, St. Petersburg, 195251 Russia}
\newcommand{\saopaulo}{Universidade de S{\~a}o Paulo, Instituto de F\'{\i}sica, Caixa Postal 66318, S{\~a}o Paulo CEP05315-970, Brazil}
\newcommand{\seoulnat}{Department of Physics and Astronomy, Seoul National University, Seoul 151-742, Korea}
\newcommand{\stonybrkc}{Chemistry Department, Stony Brook University, SUNY, Stony Brook, New York 11794-3400, USA}
\newcommand{\stonycrkp}{Department of Physics and Astronomy, Stony Brook University, SUNY, Stony Brook, New York 11794-3800, USA}
\newcommand{\subatech}{SUBATECH (Ecole des Mines de Nantes, CNRS-IN2P3, Universit{\'e} de Nantes) BP 20722 - 44307, Nantes, France}
\newcommand{\tenn}{University of Tennessee, Knoxville, Tennessee 37996, USA}
\newcommand{\titech}{Department of Physics, Tokyo Institute of Technology, Oh-okayama, Meguro, Tokyo 152-8551, Japan}
\newcommand{\tsukuba}{Institute of Physics, University of Tsukuba, Tsukuba, Ibaraki 305, Japan}
\newcommand{\vandy}{Vanderbilt University, Nashville, Tennessee 37235, USA}
\newcommand{\waseda}{Waseda University, Advanced Research Institute for Science and Engineering, 17 Kikui-cho, Shinjuku-ku, Tokyo 162-0044, Japan}
\newcommand{\weizmann}{Weizmann Institute, Rehovot 76100, Israel}
\newcommand{\wigner}{Institute for Particle and Nuclear Physics, Wigner Research Centre for Physics, Hungarian Academy of Sciences (Wigner RCP, RMKI) H-1525 Budapest 114, POBox 49, Budapest, Hungary}
\newcommand{\yonsei}{Yonsei University, IPAP, Seoul 120-749, Korea}
\newcommand{\zagreb}{University of Zagreb, Faculty of Science, Department of Physics, Bijeni\v{c}ka 32, HR-10002 Zagreb, Croatia}
\affiliation{\abilene}
\affiliation{\augie}
\affiliation{\banaras}
\affiliation{\barc}
\affiliation{\baruch}
\affiliation{\bnlcoll}
\affiliation{\bnlphys}
\affiliation{\caucr}
\affiliation{\charlesczech}
\affiliation{\chonbuk}
\affiliation{\ciae}
\affiliation{\cns}
\affiliation{\colorado}
\affiliation{\columbia}
\affiliation{\czechtech}
\affiliation{\dapnia}
\affiliation{\debrecen}
\affiliation{\elte}
\affiliation{\ewha}
\affiliation{\fit}
\affiliation{\fsu}
\affiliation{\gsu}
\affiliation{\hanyang}
\affiliation{\hiroshima}
\affiliation{\howard}
\affiliation{\ihepprot}
\affiliation{\illuiuc}
\affiliation{\inrras}
\affiliation{\instpasczech}
\affiliation{\isu}
\affiliation{\jaea}
\affiliation{\jinrdubna}
\affiliation{\jyvaskyla}
\affiliation{\kek}
\affiliation{\korea}
\affiliation{\kurchatov}
\affiliation{\kyoto}
\affiliation{\labllr}
\affiliation{\lahorelums}
\affiliation{\lawllnl}
\affiliation{\losalamos}
\affiliation{\lpc}
\affiliation{\lund}
\affiliation{\maryland}
\affiliation{\mass}
\affiliation{\michigan}
\affiliation{\muenster}
\affiliation{\muhlenberg}
\affiliation{\myongji}
\affiliation{\nagasaki}
\affiliation{\newmex}
\affiliation{\nmsu}
\affiliation{\ohio}
\affiliation{\ornl}
\affiliation{\orsay}
\affiliation{\peking}
\affiliation{\pnpi}
\affiliation{\riken}
\affiliation{\rikjrbrc}
\affiliation{\rikkyo}
\affiliation{\saispbstu}
\affiliation{\saopaulo}
\affiliation{\seoulnat}
\affiliation{\stonybrkc}
\affiliation{\stonycrkp}
\affiliation{\subatech}
\affiliation{\tenn}
\affiliation{\titech}
\affiliation{\tsukuba}
\affiliation{\vandy}
\affiliation{\waseda}
\affiliation{\weizmann}
\affiliation{\wigner}
\affiliation{\yonsei}
\affiliation{\zagreb}
\author{A.~Adare} \affiliation{\colorado}
\author{S.~Afanasiev} \affiliation{\jinrdubna}
\author{C.~Aidala} \affiliation{\columbia} \affiliation{\losalamos} \affiliation{\mass} \affiliation{\michigan}
\author{N.N.~Ajitanand} \affiliation{\stonybrkc}
\author{Y.~Akiba} \affiliation{\riken} \affiliation{\rikjrbrc}
\author{R.~Akimoto} \affiliation{\cns}
\author{H.~Al-Bataineh} \affiliation{\nmsu}
\author{H.~Al-Ta'ani} \affiliation{\nmsu}
\author{J.~Alexander} \affiliation{\stonybrkc}
\author{M.~Alfred} \affiliation{\howard}
% ...
\collaboration{PHENIX Collaboration} \noaffiliation

\date{\today}


\begin{abstract}


Two-pion interferometry measurements are used to extract the Gaussian 
radii $R_{{\rm out}}$, $R_{{\rm side}}$, and $R_{{\rm long}}$, of the pion 
emission sources produced in Cu$+$Cu and Au$+$Au collisions at several 
beam collision energies $\sqrt{s_{_{NN}}}$ at PHENIX. The extracted 
radii, which are compared to recent STAR and ALICE data, show 
characteristic scaling patterns as a function of the initial transverse 
size $\bar{R}$ of the collision systems and the transverse mass $m_T$
of the emitted pion pairs, consistent with hydrodynamiclike expansion. 
Specific combinations of the three-dimensional radii that are sensitive to 
the medium expansion velocity and lifetime, and the pion emission time 
duration show nonmonotonic $\sqrt{s_{_{NN}}}$ dependencies. The 
nonmonotonic behaviors exhibited by these quantities point to a softening 
of the equation of state that may coincide with the critical end point in 
the phase diagram for nuclear matter.

\end{abstract}

	
\maketitle


%

Studies of the matter produced in ultrarelativistic heavy ion collisions 
provide an important avenue for mapping the phase diagram for quantum 
chromodynamics (QCD)~\cite{Itoh:1970,Shuryak:1983zb,Stephanov:1998dy}. 
Quantification of the properties of the various QCD phases, as well as 
pinpointing the location of the phase boundaries and the critical end 
point (CEP) in the plane of temperature vs. baryon chemical potential 
($T, \mu_B$), are of fundamental interest~\cite{Asakawa:1989bq}.

Lattice QCD calculations indicate that the quark-hadron transition is a 
crossover at small $\mu_{B}$ or high collision energies 
(\sqsn)~\cite{Aoki:2006we,Bhattacharya:2014ara}. Experimental results from 
the Relativistic Heavy Ion Collider (RHIC) at \sqsn = 200~GeV and the 
Large Hadron Collider (LHC) at \sqsn = 2.76~TeV, indicate that this 
transition results in the production of a strongly coupled plasma of 
deconfined quarks and gluons (sQGP) with low specific shear viscosity 
$\frac{\eta}{s}$, {\em i.e.} the ratio of shear viscosity $\eta$ to 
entropy density 
$s$~\cite{Adcox:2004mh,Adams:2005dq,Aamodt:2010pa,Teaney:2003kp, 
Lacey:2006bc,Luzum:2008cw,Song:2011hk,Lacey:2010ej}.
The validation of this crossover transition 
is a necessary, albeit insufficient, requirement for the existence of 
the CEP. 

For lower beam energies or larger values of $\mu_{B}$~\cite{chemfo}, model 
calculations~\cite{Berges:1998rc,Hatta:2002sj,Stephanov:2004wx,Ejiri:2008xt} 
suggest that reaction trajectories in the ($T, \mu_{B}$)-plane could come 
close to the CEP and even cross the coexistence curve that delineates a 
first order phase transition. Thus, a current strategy for experimental 
mapping of the phase diagram is centered on beam energy scans, which 
sample reaction trajectories with the broadest possible range of $\mu_B$ 
and $T$ values.

The expansion dynamics of the matter produced in these beam energy scans 
is strongly influenced by the reaction trajectory in the ($T, 
\mu_{B}$)-plane. At the CEP or close to it, anomalies in the dynamic 
properties of the medium can drive abrupt changes in transport 
coefficients and relaxation rates to give a nonmontonic dependence of 
$\frac{\eta}{s}(T, 
\mu_B)$~\cite{Lacey:2006bc,Csernai:2006zz,Lacey:2007na}. A medium produced 
in the vicinity of the CEP could also show a stalling of the the mean 
expansion speed, $c_s$~\cite{Hung:1994eq,Rischke:1996em}, as well as a 
longer emission duration $\Delta \tau$, manifested as a difference between 
$R_{{\rm out}}$ and $R_{{\rm side}}$ ($\Delta \tau^2 \propto 
(R_{{\rm out}}^2 - 
R_{{\rm side}}^2)$)~\cite{Pratt:1984su,Hung:1994eq,Chapman:1994yv,Wiedemann:1995au,Rischke:1996em}. 
Here, the rationale is that, in the vicinity of the CEP, the equation of 
state (EOS) ``softens" considerably and this could cause the expansion to 
stall and prolong the emission duration to give 
$R_{{\rm out}}>R_{{\rm side}}$.
 
In a recent study, the acoustic scaling properties of collective flow were 
used to extract viscous coefficients as a function of 
\sqsn~\cite{Lacey:2013qua}. A striking pattern of viscous damping, 
compatible with the expected minimum of $\frac{\eta}{s}(T,\mu_B)$ for 
trajectories in the vicinity of the 
CEP~\cite{Csernai:2006zz,Lacey:2007na}, was reported. Such trajectories 
should also lead to signatures indicative of a softening of the EOS and a 
prolonged emission duration. 
Both can influence the space-time extent to give a measurable
nonmonotonic \sqsn\ dependence of the medium expansion velocity and
the duration of particle emission.

In this Letter, we use the interferometry technique of Hanbury Brown and 
Twiss (HBT)~\cite{hbt_org} to perform detailed differential measurements 
of two-pion correlation 
functions~\cite{Zajc:1984vb,Pratt:1984su,Ganz:1998zj,Adler:2004rq,Adams:2004yc,Lisa:2005dd, 
Afanasiev:2007kk,Aamodt:2011mr,Adare:2014vax} in Cu$+$Cu collisions at 
$\sqrt{ s_{{\rm NN}}} = 200$~GeV and Au$+$Au collisions at \sqsn =39.0, 
62.4, and 200 GeV. The correlation functions are then used to extract and 
compare the emission source radii to similar measurements for Au$+$Au 
collisions (\sqsn = $7-200$~GeV) by the STAR 
collaboration~\cite{Adamczyk:2014mxp} and Pb$+$Pb collisions (\sqsn = 
2.76~TeV) by the ALICE collaboration at the LHC~\cite{Aamodt:2011mr}. 
We test the scaling properties of specific combinations of the 3D radii 
that are sensitive to the expansion velocity (and thus also the speed of 
sound) and the emission time duration, which allow a search for their 
possible nonmonotonic dependence on \sqsn.


The present analysis uses the data recorded by the PHENIX experiment for 
Cu$+$Cu collisions during the 2005 RHIC running period and for Au$+$Au 
collisions during the 2007 and 2010 running periods. Event triggering, as 
well as determination of the collision vertex $z$ (along the beam axis) 
was obtained with the Beam-Beam Counters located on either side of the 
interaction point of PHENIX. This vertex was constrained to $|z|< 30$~cm. 
the charge distribution measured in the beam-beam counters, which span the 
pseudorapidity range 3.0$<$$|\eta|$$<$3.9~\cite{Adare:2013esx}. Track and 
momentum reconstruction for charged particles were performed by combining 
hits from the drift chambers (DC) and pad chambers in the PHENIX central 
spectrometers ($|\eta| < 0.35$). Charged pions were identified by 
combining time-of-flight from the two time-of-flight detectors, as well as 
the electromagnetic calorimeters~\cite{EMC}, with momentum reconstructed 
from the DC and pad-chamber hits in the magnetic field. Particles within 2 
standard deviations of the peak for charged pions in the squared mass 
distribution were identified as pions for momenta up to $\sim$ 1 GeV/$c$ 
as detailed in Ref.~\cite{Adare:2013esx}.


The two-pion correlation function is defined as the ratio 
$C_2\left({\bf q}\right)=A\left({\bf q}\right)/B\left({\bf q}\right)$, 
where $A\left({\bf q}\right)$ is the measured distribution of the relative 
momentum difference ${\bf q}={\bf p}_2-{\bf p}_1$ between particle pairs 
with momenta ${\bf p}_1$ and ${\bf p}_2$; $B\left({\bf q}\right)$ is the 
uncorrelated distribution, obtained from particle pairs in which each 
particle is selected from a different event but with similar event 
centralities, vertex positions, and charge sign. The effects of track 
merging and track splitting~\cite{Adler:2004rq,Afanasiev:2007kk} were 
suppressed via pair selection cuts in the DC and the electromagnetic 
calorimeters, as detailed in Ref.~\cite{Afanasiev:2007kk}. The relative 
momentum ${\bf q}$ is calculated in the longitudinally co-moving system, 
where the longitudinal pair momentum (along the beam direction) is zero. 
It is also decomposed into its three components, ${q}_{{\rm out}}$, 
${q}_{{\rm side}}$, and ${q}_{{\rm long}}$, following the Bertsch--Pratt 
convention~\cite{Bertsch:1989vn,Pratt:1986cc}, {\em i.e.} the ``out'' axis 
points along the pair transverse momentum, the ``side'' axis is 
perpendicular to the out axis, and the ``long'' axis points along the 
beam.

\begin{figure}[t]
\includegraphics[width=1.0\linewidth]{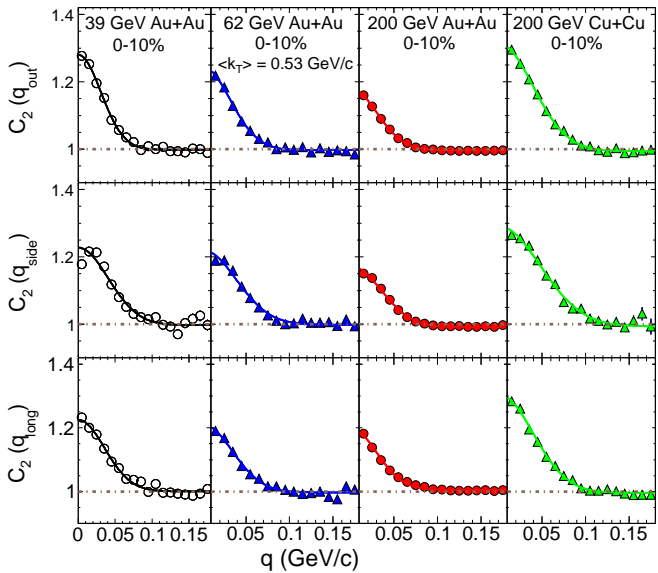}
\caption{(Color online) 
Slices of the 3D two-pion ($\pi^{+}\pi^{+}$ and $\pi^{-}\pi^{-}$) 
correlation functions for 0\%--10\% central Au$+$Au (left panels) and 
Cu$+$Cu (rightmost panel)  collisions for $\left< k_T \right> = 
0.53$~GeV/$c$ and for several beam collision energies as indicated. The 
curves represent fits to the correlation function (see text).
} 
\label{fig1}
\end{figure}

Correlation functions were studied as a function of collision centrality, 
as well as for different pion-pair transverse momenta 
$k_T = |{\bf p}_{T,1}+{\bf p}_{T,2}|/2$ or transverse mass 
$m_{T}=\sqrt{(k_{T}^2+m_{\pi}^2)}$, where $m_{\pi}$ is the pion mass.
Figure~\ref{fig1} shows a representative set of plots from the 
three-dimensional two-pion correlation functions for central (0\%--10\%) 
Au$+$Au and Cu$+$Cu collisions for $\left< k_T \right> = 0.53$~GeV/$c$ 
for several values of \sqsn.  The plots all show the familiar 
Bose--Einstein enhancement peak at low $q$.  The larger peak widths for 
Cu$+$Cu reflects the difference in the initial geometric sizes for 
0\%--10\% central Cu$+$Cu and Au$+$Au collisions.

Correlation functions were extracted for a detailed set of centrality and 
$m_T$ cuts, to allow a study of the emission sources as a function of pair 
momentum and initial-state transverse size characterized as follows. In a 
Monte Carlo Glauber (MC-Glauber) 
calculation~\cite{glauber,Lacey:2010hw,Adare:2013nff} a subset of the 
nucleons become participants ($N_{\rm part}$) in each collision by 
undergoing an initial inelastic N+N interaction.  The transverse 
distribution of these participants in the X-Y plane has RMS widths 
$\sigma_x$ and $\sigma_y$ along its principal axes. We define $\bar{R}$, 
the characteristic initial transverse size, as 
${1}/{\bar{R}}~=~\sqrt{\left({1}/{\sigma_x^2}+{1}/{\sigma_y^2}\right)}$~\cite{Bhalerao:2005mm}. 
The $\bar{R}$ and $N_{\rm part}$ were computed as a function of collision 
centrality.  Note that, for central collisions, the initial Gaussian 
radius for the collision system $R \approx \sqrt{2}\bar{R}$. The systematic 
uncertainties for these geometric quantities, obtained via variation of 
the model parameters, are less than 10\%~\cite{Adare:2013nff}.

The correlation functions were fitted with the following expression (in 
which cross-terms are assumed to be negligible) which accounts for the 
Bose--Einstein enhancement and the Coulomb interaction between pion 
pairs~\cite{Bowler,Sinyukov:1998fc}:
\begin{eqnarray}
C_{2}({\bf q}) = N [ ( \lambda (1+G({\bf q})) ) F_{c} + (1-\lambda)], \nonumber \\ 
G({\bf q}) \cong \exp( -R_{{\rm side}}^{2}q_{{\rm side}}^{2} -R_{{\rm out}}^{2}q_{{\rm out}}^{2} -R_{{\rm long}}^{2}q_{{\rm long}}^{2}),
\label{eq:sinyukov}
\end{eqnarray}
where $N$ is a normalization factor, $\lambda$ is the correlation 
strength, $F_{c}$ is the Coulomb correction factor~\cite{Sinyukov:1998fc} 
evaluated with the Coulomb wave function, and ${R}_{{\rm out}}$, 
${R}_{{\rm side}}$ and ${R}_{{\rm long}}$ are the Gaussian HBT radii which 
characterize the emission source.  $R_{{\rm long}}$ is related to medium 
lifetime and $(R_{{\rm out}}^2 - R_{{\rm side}}^2)$ is sensitive to 
$\Delta \tau$~\cite{Chapman:1994yv,Wiedemann:1995au}. Similarly, ($R_{{\rm 
side}}- \sqrt{2}\bar{R}$) gives an estimate for the expansion radius for 
small values of $m_T$.

Good fits to the correlation functions for the Cu$+$Cu and Au$+$Au systems 
were obtained (cf. Fig.\ref{fig1}) and cross-checked to confirm agreement 
with our earlier measurements for Au$+$Au 
collisions~\cite{Adler:2004rq,Afanasiev:2007kk,Adare:2014vri}. The fit 
parameters for $\pi^+\pi^+$ and $\pi^-\pi^-$ pairs were also found to 
agree within statistical uncertainties; the data for $\pi^+\pi^+$ and 
$\pi^-\pi^-$ were therefore combined. The systematic uncertainties for the 
fits were estimated via variations of the cuts used to generate the 
correlation functions (single track cuts, pair selection cuts and particle 
identification cuts). Typical values of the systematic uncertainties for 
$R_{{\rm out}}$, $R_{{\rm side}}$, and $R_{{\rm long}}$ are 5\% and do not 
exceed 8\%.

\begin{figure*}[t]
\includegraphics[width=0.998\linewidth]{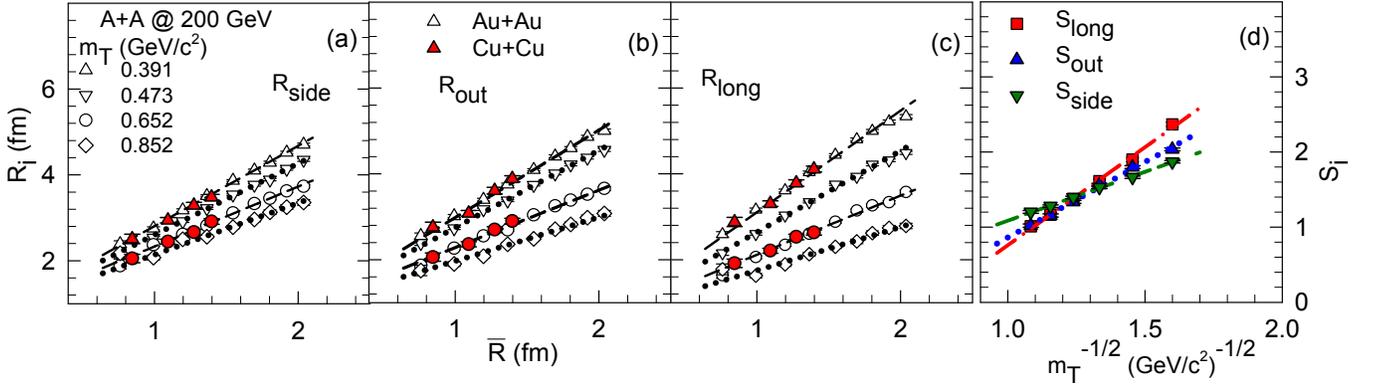}
\caption{(Color online) 
HBT radii vs. $\bar{R}$ for several $m_T$ cuts (as indicated) for (a) 
$R_{{\rm side}}$, (b) $R_{{\rm out}}$ and (c) $R_{{\rm long}}$ for 
0\%--10\%, 10\%--20\%, 20\%--30\% and 30\%--40\% Cu$+$Cu collisions, and 
0\%--5\%, 5-10\%, 10\%--15\%, 15-20\%, 20\%--30\%, 30\%--40\%, 40\%--50\%, 
50\%--60\% and 60\%--70\% Au$+$Au collisions at \sqsn =200 GeV. (d) $S_i$ 
vs. $1/\sqrt{m_T}$;  $S_i$ are slopes obtained from the respective linear 
fits to $R_{{\rm side}}$, $R_{{\rm out}}$, and $R_{{\rm long}}$ vs. 
$\bar{R}$, shown in (a), (b) and (c). The curves in (a)-(d) represent 
linear fits.
}
\label{fig2}
\end{figure*}

\begin{figure*}[t]
\includegraphics[width=0.998\linewidth]{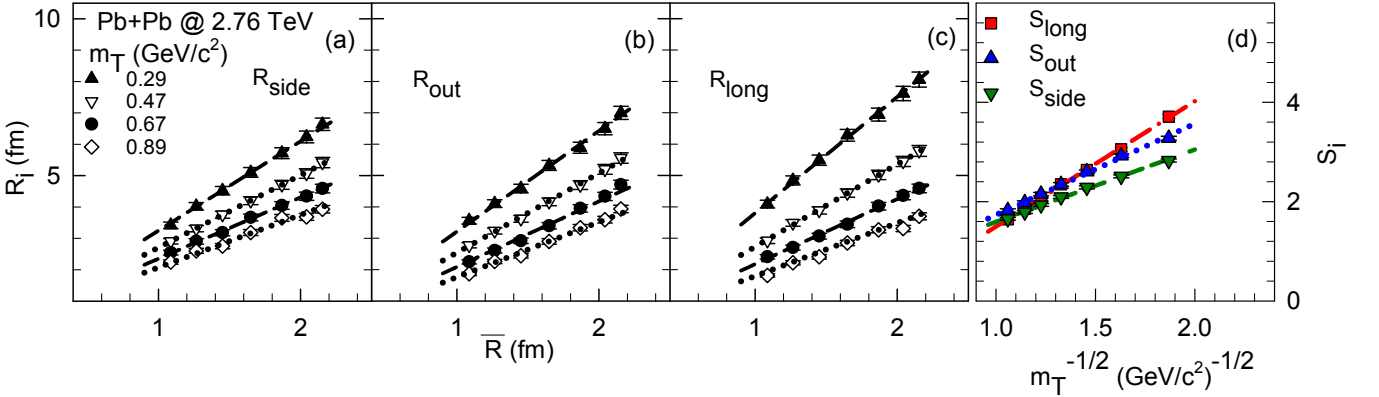}
\caption{(Color online)  
Same as Fig.~\ref{fig2}, but for Pb$+$Pb collisions at \sqsn = 2.76 TeV; 
the data are taken from Ref.~\cite{Kisiel:2011jt}.
}
\label{fig3}
\end{figure*}

At freeze-out, the space-time extent of an emission source reflects its 
initial size, its growth in size over the duration of its lifetime or 
expansion time $\tau$, as well as a diminution in size with $m_T$, due to 
position-momentum correlations. The expansion time 
$\tau\propto\bar{R}$~\cite{Lacey:2013is,Lacey:2013qua,Shuryak:2013ke}. 
Therefore, $R_{{\rm out}}$, $R_{{\rm side}}$, and $R_{{\rm long}}$ might 
be expected to scale with $\bar{R}$ for a given $m_T$. Position-momentum 
correlations reduce the magnitude of these 
radii~\cite{Adler:2004rq,Adams:2004yc,Aamodt:2011mr}, so it is instructive 
to investigate $m_T$ scaling as well.

Figure~\ref{fig2} gives a summary of the detailed centrality and $m_T$ 
dependence of the extracted radii for Cu$+$Cu and Au$+$Au collisions at 
\sqsn = 200 GeV.  Figures \ref{fig2}(a), (b) and (c)  validate the 
expected linear dependence of $R_{{\rm side}}$, $R_{{\rm out}}$ and 
$R_{{\rm long}}$ on $\bar{R}$ for both systems and show that the 
magnitudes of the radii for each system, are comparable at similar values 
of $\bar{R}$ and $m_T$ . They also indicate the expected decrease in the 
slope of the respective scaling curves (for $R_{{\rm side}}$, 
$R_{{\rm out}}$, and $R_{{\rm long}}$) with $m_T$. The latter confirms the 
important influence of position-momentum correlations which result from 
collective expansion in the Cu$+$Cu and Au$+$Au systems.  Similar scaling 
patterns were observed for the full range of \sqsn values spanned by the 
STAR data set~\cite{Adamczyk:2014mxp}.

Figures \ref{fig3}(a), (b) and (c) show that the same scaling patterns are 
also observed for the HBT radii extracted in Pb$+$Pb collisions at \sqsn = 
2.76 TeV, albeit with significantly larger magnitudes for 
$R_{{\rm side}}$, $R_{{\rm out}}$, and $R_{{\rm long}}$. Because the 
values for $\bar{R}$ in Pb$+$Pb collisions are only $\sim 5$\% larger than 
those for Au$+$Au collisions, the larger radii observed at \sqsn = 2.76 
TeV could be the result of an increase in the total system lifetime and/or 
a larger expansion velocity from RHIC to the LHC.

\begin{figure}[t] 
\includegraphics[width=1.0\linewidth]{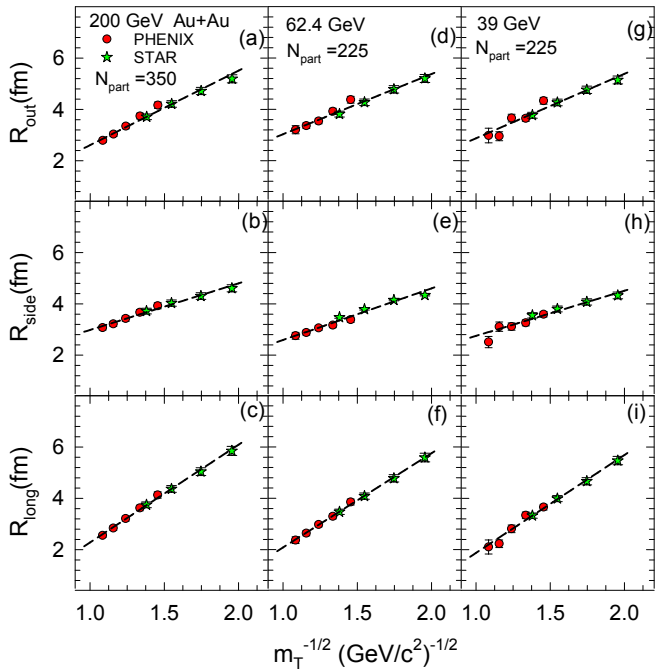} 
\caption{(Color online) 
Comparison of PHENIX and STAR HBT radii for Au$+$Au collisions at \sqsn = 
39.0, 62.4 and 200 GeV as indicated. The STAR data are taken from 
Ref.~\cite{Adamczyk:2014mxp}. The dashed curves are linear fits to the 
combined data sets. 
}
\label{fig4} \end{figure}

Linear fits were made to the plots of the HBT radii vs. $\bar{R}$ for the 
full range of $m_T$ selections [cf. dashed and dotted curves in Figs. 
\ref{fig2} and \ref{fig3}~(a)--(c)], to gain further insights on the $m_T$ 
dependence of the position-momentum correlations.  Figures~\ref{fig2}(d) and 
\ref{fig3}(d) show that the slopes $S_i$ obtained from these linear fits, 
scale as $1/\sqrt{m_T}$ and the position-momentum correlations are largest 
(smallest) in the long (side) direction. They also indicate that, for a 
given \sqsn, the full set of differential measurements for each radius, 
can be made to scale to a single curve.

Figure \ref{fig4} shows a further demonstration of these scaling patterns 
for PHENIX and STAR measurements for Au$+$Au collisions at \sqsn = 39, 
62.4 and 200 GeV. The results for two centrality or $N_{\rm part}$ 
selections indicate the characteristic $1/\sqrt{m_T}$ dependence of 
$R_{{\rm out}}$, $R_{{\rm side}}$, and $R_{{\rm long}}$ for each beam 
energy presented.  They also indicate good agreement between the PHENIX 
and STAR data sets. The PHENIX data provide a sizable extension to the 
$m_T$ reach of the available data for HBT radii at RHIC. Similar scaling 
was observed for the full range of \sqsn measurements spanned by the ALICE 
and STAR data sets~\cite{Kisiel:2011jt,Adamczyk:2014mxp}.

\begin{figure}[t]
\includegraphics[width=1.0\linewidth]{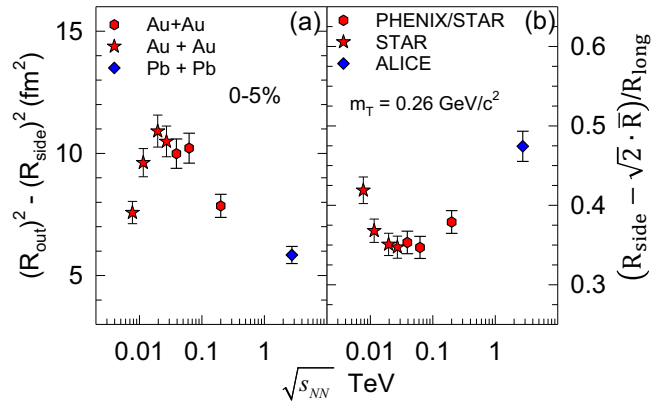}
\caption{(Color online)  
The \sqsn\ dependence of (a) $ (R_{{\rm out}}^2 - R_{{\rm side}}^2)$, 
(b) [($R_{{\rm side}}- \sqrt{2}\bar{R}$)/$R_{{\rm long}}$]. The HBT radii 
are taken from the present work and 
Refs.~\cite{Kisiel:2011jt,Adamczyk:2014mxp}. The PHENIX and STAR data 
points represent the results from fits to the $m_T$ dependence of the 
combined data sets. 
} 
\label{fig5} 
\end{figure}

The quantities $ (R_{{\rm out}}^2 - R_{{\rm side}}^2)$ and [($R_{{\rm 
side}}- \sqrt{2}\bar{R}$)/$R_{{\rm long}}$] were obtained at 
$m_T =0.26$~GeV/$c^2$ to reduce the effects of position-momentum 
correlations. These quantities, which are related to the emission duration 
and expansion velocity respectively, were investigated as a function of 
\sqsn\.   Figures~\ref{fig5}(a) and (b) show these dependencies for 
the 5\% most central collisions of the combined data sets. Similar 
patterns were observed for other $m_T$ selections, albeit with different 
magnitudes. These nonmonotonic patterns are consistent with the minimum 
observed for the \sqsn dependence of the viscous coefficients reported in 
Ref.~\cite{Lacey:2013qua}, and could be a further indication of trajectories 
passing through the softest region in the equation of state and possibly 
the CEP.

%

In summary, we have presented new PHENIX measurements of two-pion 
interferometry and used them to extract the Gaussian source radii 
$R_{{\rm out}}$, $R_{{\rm side}}$, and $R_{{\rm long}}$, of the emission 
sources produced in Cu$+$Cu and Au$+$Au collisions at several beam 
energies. The extracted HBT radii, which are compared to recent STAR and 
ALICE data, exhibit characteristic scaling patterns as a function of $m_T$ 
and $\bar{R}$ which allow an investigation of the \sqsn dependence of the 
quantities $R^2_{\rm out}-R^2_{\rm side}$ and $R_{\rm 
side}-\sqrt{2}\bar{R}/R_{\rm long}$ which are sensitive to the emission 
duration and expansion velocity, respectively. Non-monotonic dependencies 
observed in these variables may be linked to trajectories that spend a 
significant fraction of time near the softest point in the equation of 
state and possibly the CEP. Further detailed studies are required to make 
a more precise mapping, as well as to confirm that the observed patterns 
are linked to trajectories close to the critical end point in the phase 
diagram for nuclear matter.




We thank the staff of the Collider-Accelerator and Physics
Departments at Brookhaven National Laboratory and the staff of
the other PHENIX participating institutions for their vital
contributions.  We acknowledge support from the 
Office of Nuclear Physics in the
Office of Science of the Department of Energy,
the National Science Foundation, 
Abilene Christian University Research Council, 
Research Foundation of SUNY, and
Dean of the College of Arts and Sciences, Vanderbilt University 
(U.S.A),
Ministry of Education, Culture, Sports, Science, and Technology
and the Japan Society for the Promotion of Science (Japan),
Conselho Nacional de Desenvolvimento Cient\'{\i}fico e
Tecnol{\'o}gico and Funda\c c{\~a}o de Amparo {\`a} Pesquisa do
Estado de S{\~a}o Paulo (Brazil),
Natural Science Foundation of China (P.~R.~China),
Ministry of Science, Education, and Sports (Croatia),
Ministry of Education, Youth and Sports (Czech Republic),
Centre National de la Recherche Scientifique, Commissariat
{\`a} l'{\'E}nergie Atomique, and Institut National de Physique
Nucl{\'e}aire et de Physique des Particules (France),
Bundesministerium f\"ur Bildung und Forschung, Deutscher
Akademischer Austausch Dienst, and Alexander von Humboldt Stiftung (Germany),
OTKA NK 101 428 grant and the Ch. Simonyi Fund (Hungary),
Department of Atomic Energy and Department of Science and Technology (India), 
Israel Science Foundation (Israel), 
Basic Science Research Program through NRF of the Ministry of Education (Korea),
Physics Department, Lahore University of Management Sciences (Pakistan),
Ministry of Education and Science, Russian Academy of Sciences,
Federal Agency of Atomic Energy (Russia),
VR and Wallenberg Foundation (Sweden), 
the U.S. Civilian Research and Development Foundation for the
Independent States of the Former Soviet Union, 
the Hungarian American Enterprise Scholarship Fund,
and the US-Israel Binational Science Foundation.




\end{document}